\documentclass[3p,times]{elsarticle}

\usepackage{ecrc}

\usepackage{epstopdf}

\volume{00}

\firstpage{1}

\journalname{Nuclear Physics A}

\runauth{Patrick Huck}

\jid{nupha}

\jnltitlelogo{Nuclear Physics A}

\usepackage{graphicx}
\usepackage{amsmath,amssymb}
\usepackage{amssymb}

\usepackage{wrapfig}

\usepackage{lineno}

\biboptions{sort&compress}

\begin{document}

\begin{frontmatter}



\title{Beam Energy Dependence of Dielectron Production in Au+Au Collisions from STAR at RHIC}

\author{Patrick Huck (for the STAR\fnref{col1} Collaboration)}
\fntext[col1] {A list of members of the STAR Collaboration and acknowledgements
  can be found at the end of this issue.}
\address{
  Central China Normal University (HZNU), Wuhan 430079, China\\[-0.6ex]
  Goethe University, 60438 Frankfurt am Main, Germany\\[-0.5ex]
  Lawrence Berkeley National Laboratory, Berkeley, California 94720\\[-11ex]
}

\begin{abstract}
We present the energy-dependent study of dielectron production in 0-80\%
minimum-bias Au+Au collisions at $\sqrt{s_{NN}}$ of 19.6, 27, 39, and 62.4 GeV
in STAR.  Invariant mass ($M_{ee}$) and transverse momentum ($p_T$)
differential measurements of dielectron yields are compared to cocktail
simulations of known hadronic sources and semi-leptonic charmed decays. The
enhancement (excess yield) prominent in the Low-Mass Region (LMR) over the
cocktail at all energies, is further compared to calculations of $\rho$
in-medium modifications. Within statistical and systematic uncertainties, we
find that the model consistently describes this enhancement from SPS up to top
RHIC energies in its $M_{ee}$- as well as $p_T$-dependence. Dielectron
measurements drive the statistics for the future BES Phase-II program, which promises
to improve our understanding of the LMR enhancement's trend with total baryon
densities.
\end{abstract}


\end{frontmatter}


\section{Introduction}

Ultra-relativistic nucleus-nucleus collisions allow for the study of strongly
interacting nuclear matter and the associated phase diagram of the underlying
theory called Quantum-Chromo-Dynamics (QCD). They are currently the only means
to recreate nuclear matter in the laboratory as it existed during the early
universe. Under high temperatures and energy densities, nuclear matter is
predicted to undergo a phase transition into the so-called Quark-Gluon-Plasma
(QGP) in which quarks and gluons constitute the relevant degrees of freedom.
However, amongst others the following three fundamental questions need to be
raised and answered in the context of heavy-ion collisions (HICs) to prove the
creation of such a hot and dense medium~\cite{Rapp2009,Kolb2003,Adams2005}.
First, whether thermal equilibrium is reached through sufficient rescattering
during the initial stage of the collision.  Second, whether a distinctive
footprint for individual partons can be identified. Conclusive measurements of
QGP radiation would also help to resolve the third question for signals of
chiral symmetry restoration while at the same time looking for thermal
radiation from a deconfined medium, both of which are expected from QCD.\\
\indent To access manifestations of these transitions, one is left with the
spectroscopy of a quickly expanding and cooling fireball via short-lived
resonances such as the $\rho$-meson as opposed to the long-lived $\pi$- and
$\eta$-mesons decaying after freeze-out.  Modifications of the respective
spectral functions by the medium survive the fireball evolution through the
mesons' decays into $\mathrm{e}^+\mathrm{e}^-$ pairs (dielectrons) as they
serve as electromagnetic probes with negligible final-state interactions.  Note
that such in-medium information would be lost in hadronic decay channels like
$\Delta\to\pi\mathrm{N}$ and $\rho\to\pi\pi$. In addition to hadron gas and
freeze-out phases, dielectrons emanate from the initial hard scattering and
from the QGP phase via electromagnetic radiation.  Hence, dielectrons
can be considered bulk-penetrating probes by providing dynamic and direct
information about the HIC stages they originate from encoded in their invariant
mass ($M_{ee}$) and transverse momentum ($p_T$).\\
\indent The Low-Mass Region (LMR, $M_{ee}<1.1~\mathrm{GeV}/c^2$) of dielectron spectra,
on the one hand, provides information about in-medium modifications of the
vector meson's properties. The most distinct features of electromagnetic radiation
from the hot hadronic phase are unfortunately hidden by the according hadronic freeze-out
decays~\cite{Rapp2009}. However, model calculations still predict
about 50\% reduction in the $\rho/\omega$ region and about a factor of two
enhancement at $0.5~\mathrm{GeV}/c^2$ comparing a broadened to a vacuum-like
spectral function for the dielectron spectrum from direct decays of $\rho$,
$\omega$ and $\phi$ over the fireball evolution.  The $\omega/\phi$ resonances
appear to be less susceptible to the different scenarios. These modifications
might be connected to chiral symmetry restoration but the accuracy required to
test the respective differences from various scenarios in the ($\rho$) spectral
function, is experimentally very challenging.  The Intermediate-Mass Region
(IMR, $1.1<M_{ee}<3~\mathrm{GeV}/c^2$), on the other hand, can provide access
to the initial QGP temperature as well as a possibly medium-modified correlated
charm continuum.\\
CERES/NA45~\cite{CERESNA45}
first reported an enhancement in the LMR dielectron yield, and the di-muon
measurements of NA60~\cite{Arnaldi2006} established $\rho$-meson broadening
driven by baryonic interactions in the hot hadronic phase as the reason for the
measured excess. STAR measured dielectron spectra from Au+Au collisions at
$\sqrt{s_\mathrm{NN}}=200$~GeV as well as the $p_T$- and centrality-dependent
enhancement factor (ratio of LMR excess yield over
cocktail)~\cite{Adamczyk2014,YangButterworth2014}. The latter is also found in agreement
with in-medium broadened contributions to support the NA60 findings.\\ \indent
The completion of the Barrel Time-of-Flight detector (TOF) in 2010 has allowed
STAR to play a unique role in the study of dielectron
production~\cite{Adamczyk2014, DongGeurtsHuang2013Rapp2013} with
excellent particle identification, low material budget, full azimuthal
acceptance at mid-rapidity, and a wide momentum coverage~\cite{Adams2005Adamczyk2012}.
TOF efficiently rejects slow hadrons and provides
pure electron identification together with the energy loss measured in STAR's
Time-Projection-Chamber (TPC), which makes the two detectors the primary
subsystems employed in dielectron analyses at STAR. In particular, combined
with the first phase of the Beam Energy Scan Program (BES
Phase-I,~\cite{Aggarwal2010}), STAR presents the unprecedented opportunity to
map out a significant portion of the QCD phase diagram within a homogeneous
experimental environment by consistently combining various QGP and phase
transition signatures. In the context of dielectrons, STAR aims to look for
energy-dependent changes in the in-medium spectral function modifications and
QGP radiation.\hspace*{\fill}
\setlength{\columnsep}{8pt}%
\setlength{\intextsep}{0pt}%
\begin{wraptable}{r}{.4\textwidth}
  \vspace{-3pt}
  \centering
  \caption{Good minimum bias events recorded}
  \begin{tabular}{ccccc}
    Energy (GeV) & 19.6 & 27 & 39 & 62.4\\
    Events (M) & 35.8 & 70.4 & 130.4 & 67.3
  \end{tabular}
  \label{tab:evts}
\end{wraptable}%
\indent Table \ref{tab:evts} lists the datasets taken during BES Phase-I that provide
sufficient number of events for the statistics-hungry dielectron analyses. Note
that at 200~GeV, the statistics for the Au+Au and p+p measurements has been
significantly increased since QM12~\cite{YangButterworth2014}.\\
\indent The objective of this conference contribution is to present the latest results
of STAR's successful endeavor extending the dielectron measurements from top
RHIC energies down to the SPS energy regime, and hence, not only closing a wide
gap in the phase diagram, but also providing the first comprehensive
dataset of dielectron measurements with respect to energy dependence and
experimental environment. After a few comments in section \ref{sec:ana} on
selected analysis steps, section \ref{sec:res} discusses the results including
invariant mass spectra and new energy-dependent LMR $p_T$ spectra as well as
their comparison to updated simulations and model calculations. Section
\ref{sec:summ} summarizes and gives an outlook on the role of dielectron
measurements in the current STAR upgrades and the future second phase of the
Beam Energy Scan program (BES Phase-II).

\section{Data Analysis}
\label{sec:ana}
Many of the general analysis steps involved in a dielectron analysis have been
reported before in detail for PHENIX~\cite{Adare2010} and
STAR~\cite{Adamczyk2014}. We refrain from discussing them here and instead
choose to only highlight two of the most important steps, namely background
substraction and the simulation of hadronic cocktails.\\
\indent As argued in \cite{Adare2010}, $e^+/e^-$ are always created pair-wise
in HICs. Thus, the unlike-sign background can be constructed as the geometric
mean of the like-sign backgrounds independent of the respective primary
multiplicity distribution: $\langle\mathrm{BG}_{+-}\rangle = 2\sqrt{
  \langle\mathrm{BG}_{++}\rangle \langle\mathrm{BG}_{--}\rangle }$. In the LMR,
the like-sign same event method is used to reproduce the background
contribution from correlated sources, i.e. cross pairs from $\pi$ double
conversion. In this method, like-sign pairs within the same event are combined
and averaged. The resulting spectrum is then corrected for the acceptance
difference between like- and unlike-sign pairs. Another method is the
mixed-event technique in which electrons and positrons are combined from two
different events within the same event class defined by event vertex,
centrality, and event plane angle. The resulting spectra are normalized to the
corresponding same-event distributions and used in the mass region where
combinatorial contributions account for the uncorrelated background.
Particularly the IMR requires the large statistics provided by event-mixing to
obtain the necessary accuracy. The $\rho/\omega$ region of dielectron spectra
measured in the BES Phase-I energy regime exhibits a signal-to-background ratio
of about 1/100 - 1/250. Hence, the accurate subtraction of yield from
background sources is crucial to allow for the comparison of dielectron yield
from physical sources to simulations and model calculations.\\
\indent For the comparison to dielectron signal spectra, cocktails of known
hadronic sources from meson decays after freeze-out are simulated. Most
important is the input chosen for the hadrons before decaying them according to
their direct and Dalitz channels. Pseudorapidity and azimuthal distributions are
simulated flat in $\eta\in[-1,1]$ and $\phi\in[0,2\pi]$, respectively.
Transverse momentum distributions are obtained from Tsallis-Blast-Wave fits to
the latest STAR BES Phase-I data using meson-to-pion ratios from SPS with the
according STAR $\pi$ invariant yields~\cite{Tang2009,Kumar2014}. For the
Dalitz decay channels of $\pi^0$, $\eta$ and $\omega$, the Kroll-Wada formalism
is employed with the corresponding form factors taken from
PDG~\cite{Beringer2012}. Our cocktail simulations also include contributions
from correlated charmed decays of $D$- and $\Lambda_c$-mesons simulated via
PYTHIA with estimated nucleon-nucleon cross sections and scaled to Au+Au using
the average number of binary collisions at the respective energy. The decay
channels taken into account along with a representative cocktail at 19.6 GeV
are shown in figure \ref{fig:stack_ptspecLMR}.

\section{Results}
\label{sec:res}
\begin{figure}
  \centering
  \includegraphics[width = .44\textwidth]{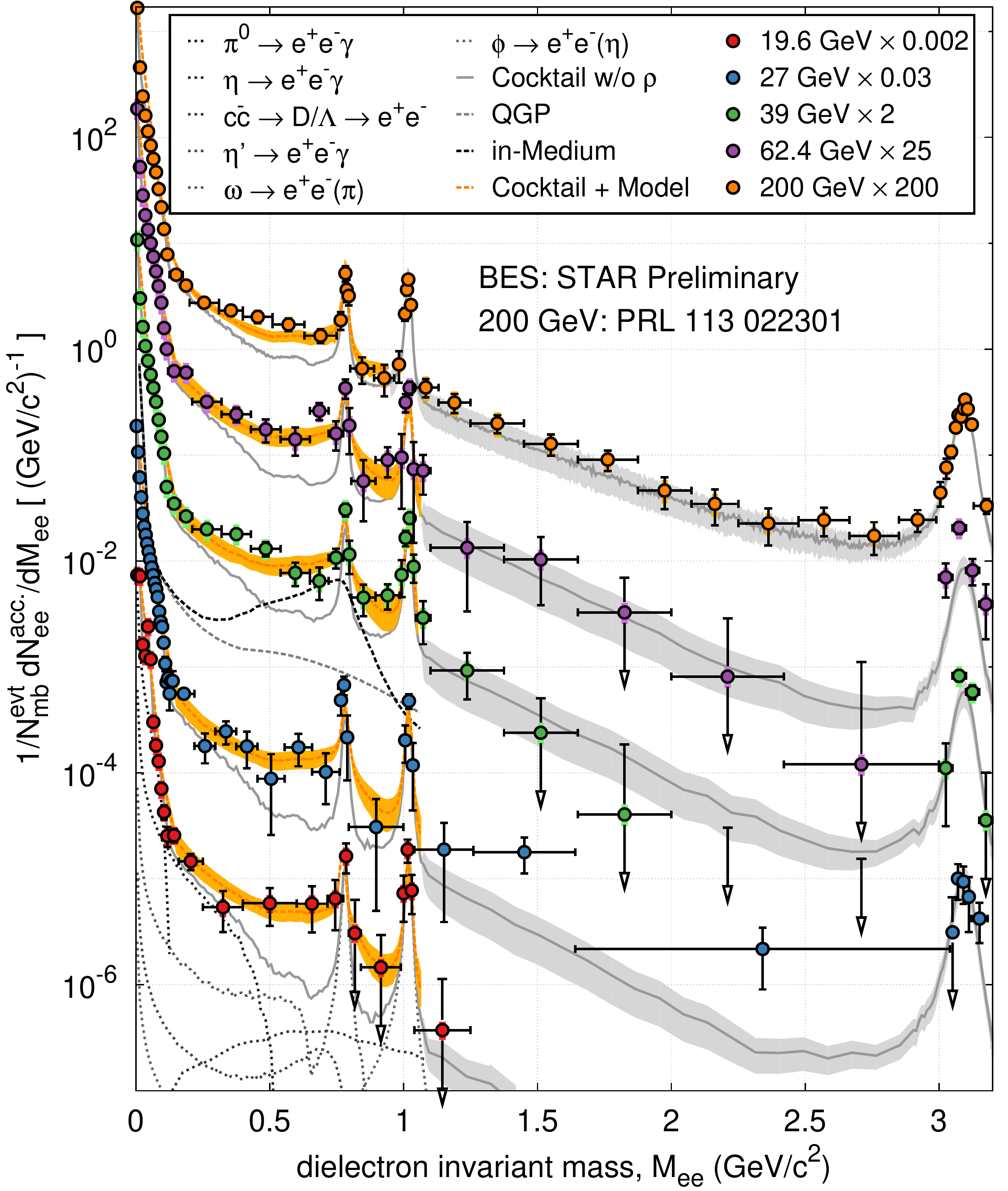}%
  \includegraphics[
  trim = 15mm 0mm 0mm 0mm, clip, width = .38\textwidth
  ]{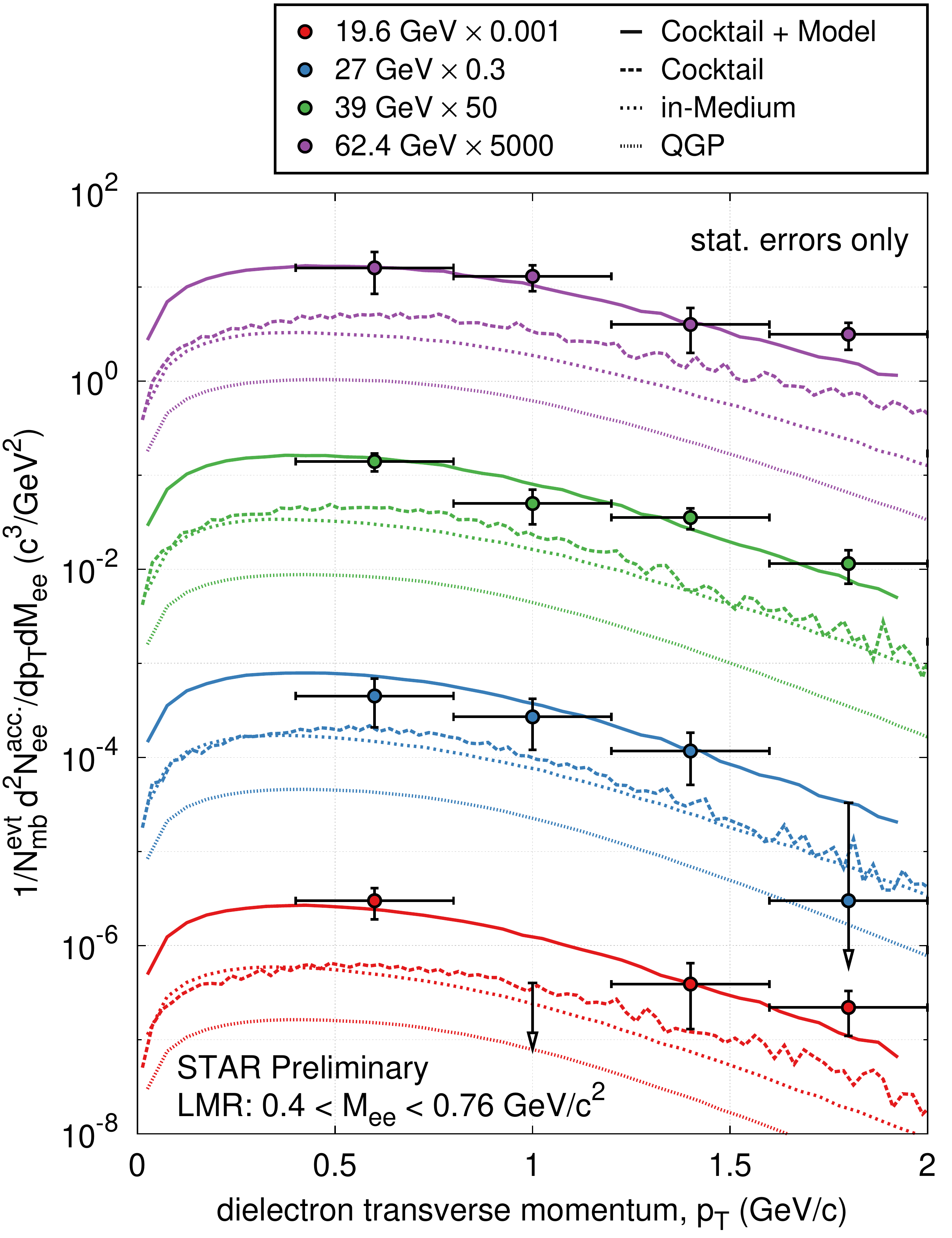}\\[-2ex]%
  \caption{
    Measured invariant mass and LMR $p_T$ dielectron spectra for BES Phase-I
    compared to cocktail simulations and model calculations~\cite{Rapp2000,
      YangButterworth2014}. (Left) In addition to the total cocktail
    (grey), the single vector meson contributions are depicted for 19~GeV. For
    39~GeV, the two model contributions from hadronic gas and QGP phase are
    shown next to the total expected yield from cocktail and model (orange).
    Also, contributions from photon conversions are not removed from the data
    at 19~GeV. (Right) Solid lines compare the $p_T$-dependence of LMR yields
    to the expectations from cocktail and model. Dashed and dotted lines
    represent the different contributions to the total yield. The error bars on
    the $p_T$-dependent data are statistical only.\protect\\[-5ex]
  }
  \label{fig:stack_ptspecLMR}
\end{figure}
\begin{wrapfigure}{r}{.4\textwidth}
  \centering
  \includegraphics*[width = .4\textwidth]{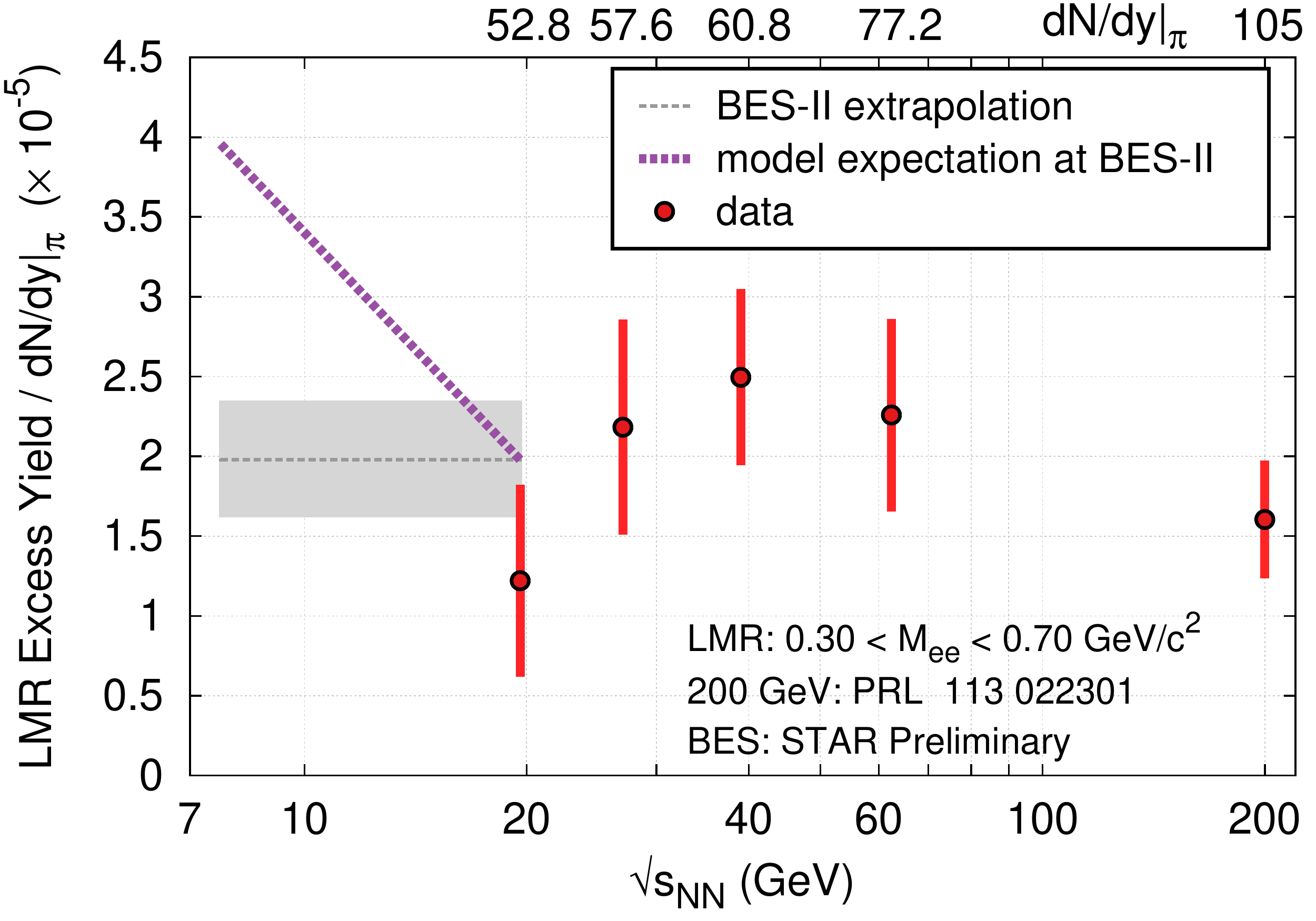}\\[-2ex]%
  \caption{
    Measured BES Phase-I LMR excess normalized to the respective
    dN/dy$\vert_{\pi}$. The grey box represents the average excess from BES
    Phase-I projected into the BES Phase-II regime with the size of the
    uncertainty taken from 200 GeV.  The magenta line depicts the expected
    trend of the LMR excess in the BES Phase-II regime as suggested by the
    according quantity in PHSD calculations.
  }
  \label{fig:excess}
\end{wrapfigure}

Figure \ref{fig:stack_ptspecLMR} left, depicts the invariant mass spectra of
all STAR dielectron measurements enabling a systematic study of dielectron
production for the wide energy range from 19.6 to 200 GeV. Cocktail simulations
generally show good agreement with the data. Only in the LMR, all spectra
exhibit a consistent excess over the cocktail. Note that contributions from
direct $\rho$ decays after freeze-out are not included in the cocktails as they do not
account for the magnitude of the enhancement~\cite{Arnaldi2009}. Instead, the
in-medium model calculations treat the full evolution of the $\rho$-meson and
hence need to be added on top of the cocktail. The comparison
of model calculations to the invariant mass dependence of dielectron LMR yields
supports the conclusion that, within systematic uncertainties, in-medium
modifications of the $\rho$ spectral function consistently describe the LMR
enhancement from SPS to top RHIC energies~\cite{Rapp2000}.
This can especially be seen in the fact that, in the LMR, electromagnetic
radiation from the hadronic gas clearly dominates over contributions from the
QGP as indicated by the model calculations. In figure \ref{fig:stack_ptspecLMR}
right, STAR's measurements are now extended to the $p_T$ dependence of
dielectron production in the LMR covering the full phase space available to
dielectrons. In line with the invariant mass dependence, we observe consistency
with the in-medium broadened spectral function scenario at all energies. This
further supports the conclusion of it being the correct description for the LMR
excess at BES Phase-I energies.\\
\indent The quality of the STAR data also allows for the systematic measurement
of LMR enhancement factors and excess yields with respect to their energy
dependence in the BES Phase-I regime. Due to the seeming CP invariance of
the strong interaction, in-medium modifications to the $\rho$ spectral function
are expected to depend on total instead of net baryon density. For energies in
BES Phase-I, the total baryon density is approximately constant which means
that an energy-dependent enhancement might be directly related to temperature
and system evolution. However, in figure \ref{fig:excess} we do not observe a
strong energy dependence in BES Phase-I.

\section{Summary \& Outlook}
\label{sec:summ}
\setlength{\columnsep}{8pt}%
\setlength{\intextsep}{0pt}%
STAR's BES program provides a unique opportunity to address long-standing
questions regarding the consequences of in-medium modifications to dielectron
spectra. The measurements provide the first comprehensive dataset to serve the
better understanding of the LMR enhancement regarding its $M_{ee}$, $p_T$ and energy
dependence. We observe that the LMR excess at all BES Phase-I energies is
consistently in agreement with in-medium modifications to the $\rho$ spectral
function. The corresponding yields do not show a strong energy dependence due
to an approximately constant total baryon density. For the energy regime below
20 GeV, measurements of the total baryon density as well as $\rho$-meson based
PHSD calculations suggest an increase of about a factor of two in LMR
excess~\cite{Cassing2000Linnyk2014}. High-statistics measurements in BES
Phase-II should provide enough accuracy to study these predictions and further
strengthen our understanding of the LMR enhancement and its
origin~\cite{SN0598}. Besides enhanced statistics, BES Phase-II will provide
improved tracking due to the proposed iTPC upgrade, and improved capabilities
for the measurement of dimuons, for instance. Moreover, the recently completed
upgrades of the HFT and MTD detector subsystems enable the study of a possibly
medium-modified charm continuum and QGP radiation. This is especially important
for the LMR since the $c\bar{c}$ contribution to the total cocktail in the 0.4
- 0.7 GeV/$c^2$ mass region increases from about 20\% at 19.6 GeV up to about
60\% at 200 GeV.\\
\indent In conclusion, STAR's measurements during BES Phase-I provide
high-quality datasets essential to address current challenges in dielectron
physics and STAR will continue along this path during BES Phase-II.

\renewcommand{\refname}{\vskip -8mm}
\bibliographystyle{elsarticle-num}
\bibliography{QM14ProcRefsNoTitle}

\end{document}